# Global optimization in the discrete and variable-dimension conformational space: The case of crystal with the strongest atomic cohesion

Guanjian Cheng[a,b], Xin-Gao Gong[c,b], and Wan-Jian Yin[a,b,c,*]

[a]College of Energy, Soochow Institute for Energy and Materials InnovationS (SIEMIS), and Jiangsu Provincial Key Laboratory for Advanced Carbon Materials and Wearable Energy Technologies, Soochow University, Suzhou, 215006, China

[b]Shanghai Qi Zhi Institute, Shanghai, 200232, China

[c]Key Laboratory for Computational Physical Sciences (MOE), Institute of Computational Physical Sciences, Fudan University, Shanghai, 200438, China

[d]Light Industry Institute of Electrochemical Power Sources, Soochow University, Suzhou, 215006, China

We introduce a computational method to optimize target physical properties in the full configuration space regarding atomic composition, chemical stoichiometry, and crystal structure. The approach combines the universal potential of the crystal graph neural network and Bayesian optimization. The proposed approach effectively obtains the crystal structure with the strongest atomic cohesion from all possible crystals. Several new crystals with high atomic cohesion are identified and confirmed by density functional theory for thermodynamic and dynamic stability. Our method introduces a novel approach to inverse materials design with additional functional properties for practical applications.

**optimization | inverse design | universal machine learning potential | atomic cohesion**

**Significance Statement**
Energy optimization represents fundamental physics in diverse aspects of scientific disciplines. Conventional optimizations were performed either in continuous space (continuous optimization) or in discrete space (combinational optimization). This study introduces an approach for the first time to simultaneously do the discrete/combinational and continuous optimization, and solves a fundamental and challenging problem since 1960s, regarding which crystal has the largest atomic cohesion. It presents a pivotal advance that can extend the optimization to wide and practical fields, such as inverse design of materials.

A long-standing goal in condensed matter and materials science is to discover materials with targeted physical properties $P$ without any constraints on atomic composition (denoted as $a$), chemical stoichiometry (denoted as $c$), and crystal structures (denoted as $s$), which mathematically relies on globally optimizing $P$ on the full configuration space of $(a, c, s)$, according to $P = f(a, c, s)$ [1]. However, there are two main obstacles to global optimization. One is the distinct characters between discrete variables $a$ and $c$, and continuous variable $s$, which causes global optimization to be a combination of continuous and discrete optimizations, known as the difficult problem of combinatorial optimization [2]. The other is the variable dimension of the full $(a, c, s)$ space, where the dimension of $s$ is dependent on that of $a$ and $c$. Optimization on the variable-dimension space where the space dimension itself is a variable is an open problem in mathematics. Therefore, the traditional approaches evaluate $P$ in a very narrow configuration space with a fixed dimension via experiments or density functional theory (DFT) calculations. These evaluations include, but are not limited to, the following strategies: (*i*) searching $a$ for fixed $c$ and $s$, such as screening tetrahedral semiconductors [3], perovskite $ABX_3$ [4], elpasolite $ABC_2D_6$ [5], and spinel $AB_2C_4$ [6] by brute force atomic substitutions on the fixed prototype structure. (*ii*) Searching $s$ for fixed $a$ and $c$, such as crystal structure prediction [7–10] by evaluating $P$ of various possible structures $s$ at a given $a$ and $c$. (*iii*) Searching $c$ and $s$ for a fixed $a$, such as in the construction of a convex hull and phase diagram [11] and the design of direct-gap Si [12].

In addition to the algorithm, the insurmountable obstacle to global optimization is the astronomical number of possible material candidates in the full space. For example, considering a binary compound $X_mY_n$ (X ≠ Y), where X and Y can be any of eighty-six elements in the first six periods in the periodic table and m, n ≤ 10 in the supercell, the dimensions of the variables $a$, $c$, and $s$ are 3655, 100, and 3×($m+n+2$), respectively. In this case, the structural search is a nontrivial task; thus, hundreds of DFT structural optimizations are required to determine the ground-state structure for a 20-atom supercell system [8,13]. Moreover, the dimension of the full configuration space is the product of the three variables mentioned above, making it challenging to exhaustively enumerate and optimize the full space via the DFT approach.

Advanced efficient methods such as machine learning have been effective in custom-fitted systems [14–16]. However, generalizing these methods to a wide range of elements for unfitted or untrained systems is challenging, and thus they cannot be implemented for optimizing $a$. For example, in an artificial neural network (NN) potential, the number of descriptors, which is proportional to the number of local environments, increases tremendously with the number of elements in the systems. In particular, when the number of elements ($N_{elem}$) is equal to 40, an ideally tetrahedral crystal has at least $10^6$ times different local environments than that for $N_{elem}$ = 1 (see Supplementary Materials for details). Thus, this approach requires $10^6$ times as many descriptors and training datasets to achieve the same accuracy as that for $N_{elem}$ = 1, which is unaffordable.

In typical NN architecture, the total energy $E$ is approximated as individual embedded atoms in local environments. In contrast, a crystal graph neural network (CGNN) evaluates $E$ as a global function of ($a$, $c$, $s$) by initially converting ($a$, $c$, $s$) to a crystal graph $G$, defined as nodes (atoms) $V$ and edges (bonds) $X$. Moreover, instead of hand-crafted descriptors, elemental embedding is learned during training, and the atomic environments and interactions are included in the operations of CGNN so that no descriptor explosion problems arise, and the model demonstrates the scalability to wide elements. Thus, combined with the optimization algorithm, the CGNN can predict the crystal structure [9]. Furthermore, the recent inclusion of unstable structures has improved the CGNN model to capture the curvature and high-energy part of PES. Therefore, it may act as a universal potential (UPot), covering a wide range of elements in the periodic table [17–19].

In this letter, we show that the CGNN potential combined with Bayesian optimization (BO) can be applied to the global optimization of the full space of the atomic composition, chemical stoichiometry, and crystal structure for desired physical properties. We selected the crystal with the highest atomic cohesion as the target. Computing cohesive energies, a classical problem, has encompassed the development history of solid-state physics [20–24]. Nevertheless, the fundamental question regarding which bonding type and crystal have the largest atomic cohesion has been a challenge that has never been answered. Atomic cohesion is closely related to the melting point of a crystal; therefore, the global optimization of atomic cohesion is equivalent to the inverse design of high-performance refractory materials with applications in a wide range, from smelting furnaces and rocket heat shields to jet engine nozzles [25]. This study describes a solution to the global optimization considering the full ($a$, $c$, $s$) space by combining a UPot and a BO to implement the inverse design of the crystal with the largest atomic cohesion.

A UPot potential was trained within the framework of the CGNN with three-body interactions [17] using the DFT-calculated total energy, atomic force, and cell stress obtained from the Materials Project database. The details of the GN architecture, model training, and verification can be found in the Supplementary Materials and Ref. [9,17]. To compare the data fairly, all DFT calculations presented in this letter were performed with VASP using Python Materials Genomics (Pymatgen) and the Perdew-Burke-Ernzerhof (PBE) functional, which is consistent with the MP database parameter settings.

BO was combined with UPot to accelerate the identification of global/local minima in the full ($a$, $c$, $s$) space. BO was selected because this algorithm has shown its efficiencies for discrete optimization by selecting optimal candidates among millions for alloy material design [26,27] and for continuous optimization by crystal structure prediction [9,10]. A flowchart of the approach combining UPot and BO is shown in Fig. 1(a). A pool of elemental candidates is given in box 1 ($N_{pool}$, the number of elements in the pool), and the crystals are generated sequentially in ($a$, $c$, $s$) spaces. First, the atomic compositions are randomly chosen, with a number of compositions up to $N_{comp}$ (box 2). Then, the chemical stoichiometry for each composition is generated with a maximum $N_{chem}$ (box 3). Finally, the crystal structure is constructed based on atomic composition and chemical stoichiometry (box 4).

Herein, we set $N_{pool}$ = 42 (Fig. S3), $N_{comp}$ = 2 or 3, and $N_{chem}$ = 5. Following the above procedure (from box 1 to box 4), 200 crystals were randomly generated with the chemical formula $A_mB_n$ ($N_{comp}$ = 2) in the full ($a$, $c$, $s$) space. Fig. 1(c) shows the compositional distributions of 200 structures. Their cohesive energies were obtained by UPot using the equation $P = f(a, c, s)$ (box 5). Starting from the 201$^{st}$ step, the BO method actively learns a surrogate model of the PEL based on the existing dataset [($a$, $c$, $s$), $P$] on which it performs both exploitation and exploration (box 6). Subsequently, the approach suggests a new candidate with a balance between exploitation and exploration [28] (line I). A new data pair [($a$, $c$, $s$), $P$] is iterated to refine the surrogate model, which suggests a new candidate for the next-round of iterations. The cohesive energy rapidly converges [Fig. 1(b)], with the compositional elements quickly converging around C and Zr [Fig. 1(c)]. The UPot-BO approach only takes 878 steps (~ 3.44 CPU hours) to find the target compound, ZrC, of the rock-salt structure with the lowest atomic cohesive energy. Note that when $N_{comp}$ = 3 ($A_lB_mC_n$), $E_c$ converges to the binary compound ZrC [Fig. 1(d)]. As shown in Fig. 1(e), the compositional elements are randomly distributed in A, B, and C in the first 200 steps for initialization but rapidly converge to binary compounds with ternary compounds for exploration. In sum, we have combined a universal potential and Bayesian optimization to perform the full space optimization for inverse material design with high efficiency. As a case study, the approach identifies the compounds with the strongest atomic cohesion.

To illustrate the reliability and generalizability of the proposed UPot and UPot-BO approaches, we selected the system of ZrC with eight atoms to examine its PEL with respect to the DFT results. Four hundred structures of ZrC with eight atoms in the cell were randomly generated, and their atomic cohesion energies were evaluated by UPot and DFT calculations. Subsequently, the 30-dimensional (24 positional + 6 lattice dimensions) PEL was reduced to two dimensions for clear vision using the uniform manifold approximation and projection (UMAP) algorithm [29]. Such dimension-reduced PEL maintained the original structural topology and physics and acted as a practical approach to analyzing high-dimensional PEL [30]. As shown in Fig. 2(a), UPot can reproduce the qualitative shape of the DFT PEL [Fig. 2(b)], even though the quantitative mean absolute error (MAE) (1.27 eV/atom) is much larger than the quantum accuracy (~ 1 meV). This is understandable because quantum accuracy and universal applicability to a wide chemical space are conflicting requirements, and UPot sacrifices its accuracy to increase its chemical applicability. Note that the high accuracy most machine-learning potentials try to achieve is, in practice, not always correspond to the particular scientific question of interest [31]. For UPot, the MAE around the minimum of the basins was lower than that of the high-energy region of the basin, as shown in Fig. 2(c). These results show that UPot can capture rough features of the energy landscape at the high-energy level while retaining a good resolution around the low-energy local minima. The proposed UPot should be suitable for optimization problems where only the static structures (zero-order value of the minimum in the PEL) and their energy values matter instead of dynamic problems involving forces and high-energy regions. In this sense, the optimization problem requires less standard accuracy for UPot than the dynamic problem. Notably, the PELs in Fig. 2(a-b) are created based on the data of 400 random structures that were not part of the training dataset for UPot, ensuring a good interpretive capability of the proposed UPot to enable the description of hidden crystals during the optimization process.

The proposed UPot showed that the $s$ space of $Zr_4C_4$ [Fig. 2(a)] was extended to the ($c$, $s$) space of $Zr_mC_n$ (m, n:1-5) [Fig. 2(d)] and finally to the full ($a$, $c$, $s$) space of $A_mB_n$ [Fig. 2(e)]. The significant increase in computational cost increases the processing demand in UPot. Therefore, the number of sampling points on the $s$ space for each fixed ($a$, $c$) was reduced from 400 to 100 and 30, as shown in Fig. 2(a), Fig. 2(d), and Fig. 2(e), respectively, which required the calculations of $903 \times 25 \times 30 = 677250$ structures to generate such a full landscape. Note that this number is not affordable by direct DFT calculations. The extension of the ($a$, $c$, $s$) space to ternary $A_lB_mC_n$ [l, m, n:1-5] would require at least ~$10^8$ structures for generating a similar full landscape, as shown in Fig. 2(e). Consequently, enumerating all possible crystals to have a full landscape and identifying the optimal one is challenging. Thus, an optimization algorithm such as UPot-BO must be implemented.

It is noted in Fig. 2(e) that the crystals with high atomic cohesion are the combinations of the high-valence cations Ti/Zr and strongly electronegative anion B, C, N, and O with small size, facilitating the formation of strong ionic bond between cation and anion and strong covalent bond between anion themselves. We selected the fifty compounds with the lowest atomic cohesion identified by the UPot-BO approach within 5000 steps, as shown in the shaded area in Fig. 1(b)(d). The detailed results are listed in Table S1. Moreover, DFT was calculated to determine the structural relaxation of the crystals. The results of the UPot, DFT, and DFT-relaxed are shown in Fig. 3. The relaxation energy, defined as the energy difference between the initial and final structures during atomic relaxation, has a mean absolute error (MAE) of less than 54 meV/atom. This result indicates that UPot can describe the structural minimum, confirming the results in Fig. 1(a-b) that UPot can qualitatively describe the DFT-calculated PEL with quantitative error. Because these compounds were derived from optimization in the full ($a$, $c$, $s$) space, they are certainly minima in $s$ spaces and have stable crystal structures. We selected ten of them for different $a$ spaces, as shown in Table 1. DFT calculations of the convex hull and phonon spectra confirmed their thermodynamic and dynamical stability [see Supplementary Materials]. Five of ten known compounds that can be found in MP. The strong cohesion of ZrC (ZrN) with rock-salt structure can be understood from the strong ionic bond between $Zr^{4+}$ ($Zr^{3+}$) and $C^{4-}$ ($N^{3-}$). Interestingly, the graphite which should have the strongest covalent bond is merely a little weaker (~ 124 meV/atom) than that of ZrC. For $ZrB_2$, B-B formed a strong covalent bond with the hexagonal structure and intercalated Zr, donating electrons to the hexagonal boron sheet. As the consequence of high atomic cohesion, both $ZrB_2$ and ZrN are known as ultra-high temperature ceramics with melting points of 3246 °C and 2952 °C, respectively, and are promising for demanding aerospace and nuclear applications. Five compounds were not in ICSD and were thus discovered as new materials. The results confirmed the applicability of our approach and pointed out a way to design crystals with high atomic cohesion.

Our approach, for the first time, optimized the full space of material configurations and enabled the inverse design of materials without any constraint on atomic composition, chemical stoichiometry, or crystal structure. Although the number of atomic compositions ($N_{comp}$) and chemical stoichiometry ($N_{chem}$) in this study was set to three and five, respectively, these parameters can be flexibly tuned in the approach to expand the search space. The applicability of this approach relies on a universal graph with deep learning potential and an optimization algorithm. The universal machine-learning potential is a fast-developing research frontier [17–19,32], and any future progress would help improve the efficiency and accuracy of the current approach.

In conclusion, we combined a universal potential based on a crystal graph neural network and an optimization algorithm, enabling the global optimization of the full space of the atomic composition, chemical stoichiometry, and crystal structure for inverse material design. We demonstrated the applicability of the algorithm to identify the crystal with the largest atomic cohesion within the binary compound $A_mB_n$ and ternary compound $A_lB_mC_n$, where A, B, and C could be any of the 42 elements. Moreover, l, m, and n were considered up to 5, and the crystal structures were unconstrained. Instead of enumerating up to ~$10^8$ possible crystals with an unaffordable cost, the approach identified rock-salt ZrC as the crystal with the largest atomic cohesion by searching only $10^3$ crystals with an extremely low computational cost (~20 CPU h).

Meanwhile, the proposed approach discovered a series of new compounds with high thermodynamic and kinetic stability, which have never been reported in the literature and databases. This approach may open a new avenue for inverse material design for other targeted properties such as bandgap, bulk/shear modulus, and thermal conductivity for practical applications in functional materials.

Author Contributions: W. Y. supervised the project. X.G. and W.Y. conceived the idea. G.C. wrote the code and conducted the calculations. W.Y. wrote the first draft and all authors discussed the results and revised the manuscript.

*To whom correspondence may be addressed. Email: wjyin@suda.edu.cn

Acknowledgment
The work was supported by the National Key Research and Development Program of China (grant Nos. 2020YFB1506400), National Natural Science Foundation of China (grant Nos. 11974257, 12188101), Jiangsu Distinguished Young Talent Funding (grant No. BK20200003), Yunnan Provincial Key S&T Program (grant No. 202002AB080001-1), Soochow Municipal Laboratory for low-carbon technologies and industries. DFT calculations were carried out at the National Supercomputer Center in Tianjin [TianHe-1(A)].

TABLE I. Ten selected compounds with the lowest $E_b$'s and different atomic compositions correspond to the shaded area in Fig. 1(b) & (d).

| # | compound | Number of atoms in cell | $E_b$ (eV/atom) | structure index in MP |
|---|---|---|---|---|
| 1 | ZrC | 8 | -8.034 | mp-2795 |
| 2 | $Zr_4C_3O_1$ | 8 | -7.931 | - |
| 3 | C | 8 | -7.910 | mp-937760 |
| 4 | TiC | 8 | -7.869 | mp-631 |
| 5 | $Zr_4C_2N$ | 7 | -7.776 | - |
| 6 | $Zr_4BC_5$ | 10 | -7.748 | - |
| 7 | ZrN | 4 | -7.592 | mp-1352 |
| 8 | $Zr_4ScC_5$ | 10 | -7.510 | - |
| 9 | $ZrB_2$ | 6 | -7.402 | mp-1472 |
| 10 | $Ti_5N_4$ | 9 | -7.389 | - |

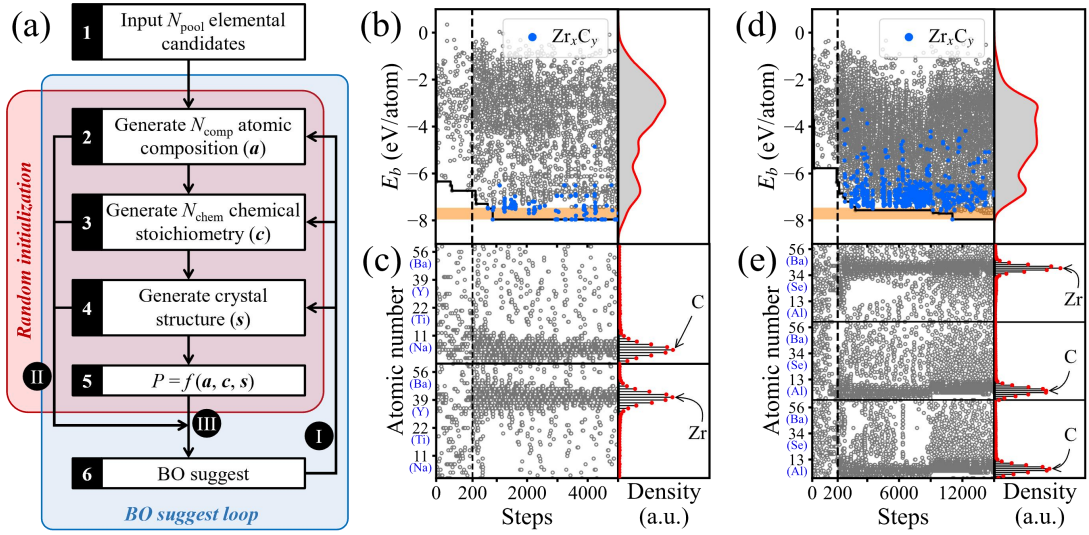

FIG. 1. (a) Flowchart of the global optimization processes. The optimization process of atomic cohesion energy $E_b$ for (b) binary compound $A_mB_n$ and (d) ternary compound $A_lB_mC_n$, with the corresponding atomic compositions shown in (c) and (e), respectively. The compounds with the chemical formula $Zr_xC_y$ (x and y are from 1 to 5) are marked as blue colors. The materials within the shaded area ($E_b < -7.40$ eV/atom) are shown in Table S1.

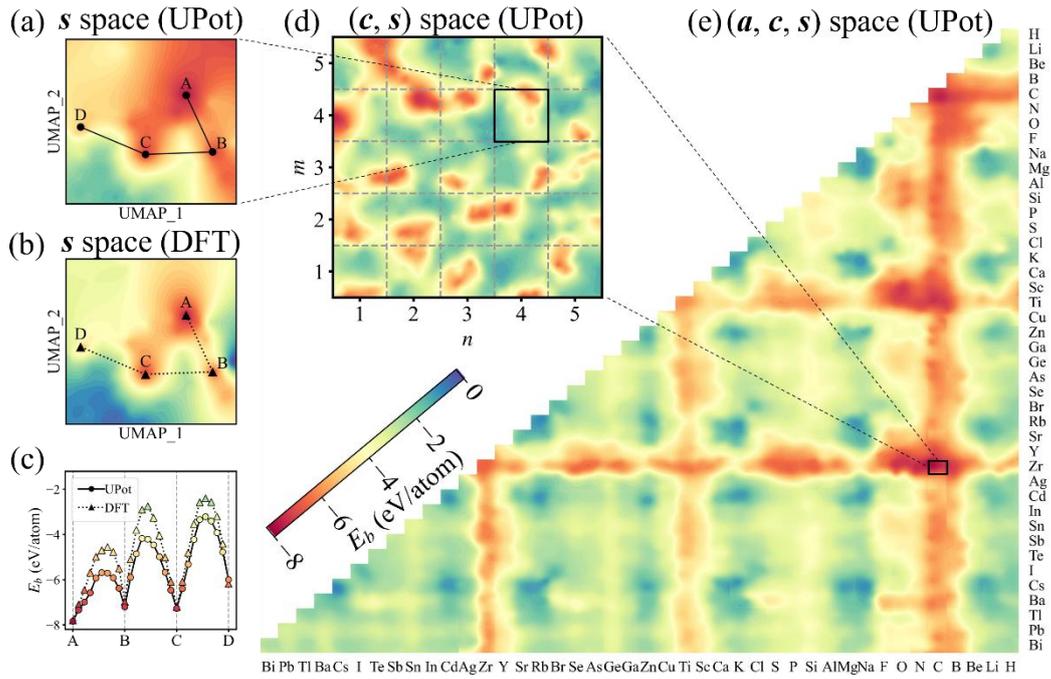

FIG. 2. The two-dimensional UMAP of PEL on $s$ space for $Zr_4C_4$ fitted from 400 structures with their $E$'s derived from (a) UPot and (b) DFT. (c) The comparison of $E$'s between UPot and DFT along the lines A-B-C-D as marked in (a) and (b). (d) The PEL on the ($c, s$) space of $Zr_mC_n$ (m, n: 1-5). (e) The PEL on the full ($a, c, s$) space of $A_mB_n$ (m, n: 1-5; A, B are any of the 42 elements in Fig. S3)

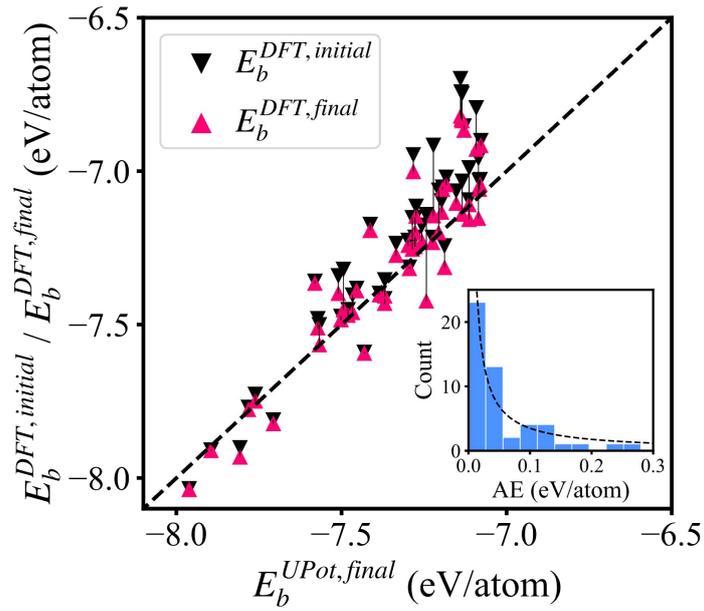

FIG. 3. Atomic cohesive energies $E_b$ of the fifty compounds in the shaded area in Fig. 1(b) & (d) derived from UPot ($E_b^{UPot}$) and DFT ($E_b^{DFT, initial}$). Atomic relaxations were performed on the DFT level, and the $E_b$ in the relaxed structure was recorded as $E_b^{DFT, final}$. The detailed values are shown in Table S1.

# Supplementary Information for

# Global optimization in the discrete and variable-dimension space of atomic composition, chemical stoichiometry, and crystal structure: The case of crystal with the strongest atomic cohesion

Guanjian Cheng[a,b], Xin-Gao Gong[c,b], and Wan-Jian Yin[a,b,c,*]

[a]College of Energy, Soochow Institute for Energy and Materials InnovationS (SIEMIS), and Jiangsu Provincial Key Laboratory for Advanced Carbon Materials and Wearable Energy Technologies, Soochow University, Suzhou, 215006, China

[b]Shanghai Qi Zhi Institute, Shanghai, 200232, China

[c]Key Laboratory for Computational Physical Sciences (MOE), Institute of Computational Physical Sciences, Fudan University, Shanghai, 200438, China

[d]Light Industry Institute of Electrochemical Power Sources, Soochow University, Suzhou, 215006, China

**Training the UPot**

In M3GNet, the crystal graph is represented by $G(\{v_i\}, \{e_{ij}\}, \{x_i\}, M, u)$, where $v_i$ is the elemental attribute of $i$th atom, $e_{ij}$ is the bond attribute between $i$th and $j$th atoms, $x_i$ is the coordinates of $i$th atom, $M$ is the 3×3 lattice matrix in the crystal, and $u$ is a global attribute, such as temperature and pressure. The details of the GN architecture and model training and verification can be found in Ref. [1,3]. We considered materials containing elements within the first five periods in the periodic table because DFT results based on local density approximation or grand gradient approximation in the training set may not effectively deal with elements with $f$ electrons in the sixth and seventh periods. We focused on nonmagnetic materials and excluded $d$ elements that often exhibit partially occupied $d$ orbitals because the spin states and related magnetic orders may introduce uncertainties to the results; thus, they required a particular treatment [2]. Rare gas and radioactive elements were also excluded, as they are rarely considered in typical research. Therefore, we considered forty-two elements, as shown in the Fig. S3. The dataset contained 58,864 items and covered 19,716 compounds. The data contained the total energies, forces, and stresses. Moreover, the first, middle, and last ionic steps were selected in the DFT structure relaxations for each compound. The dataset was divided into a training set (90%), a validation set (5%), and a test set (5%). The final performance of M3GNet on the test set was MAE (energy) = 14.27 meV/atom, MAE (force) = 103.65 meV/Å, and MAE (stress) = 0.23 GPa, as shown in Fig. S1. To compare the data fairly, all DFT calculations presented in this letter were performed with VASP using Python Materials Genomics (Pymatgen) and the Perdew-Burke-Ernzerhof (PBE) functional, which is consistent with the MP database parameter settings.

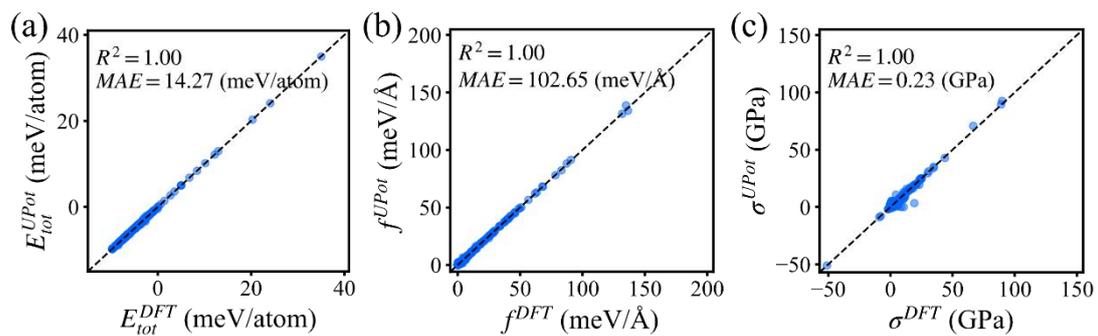

Fig. S1. UPot-predicted and DFT-calculated (a) total energies, (b) atomic forces, and (c) bulk modulus for 2,944 testing materials.

Table S1. 50 compounds with the lowest $E_b$'s and different atomic compositions correspond to the shaded area in Fig. 1(b)(d).

| # | compound | $E_b^{UPot,final}$ | $E_b^{DFT,inital}$ | $E_b^{DFT,final}$ |
|---|---|---|---|---|
| 1 | Zr4C4 | -7.961 | -8.035 | -8.036 |
| 2 | Zr4C3O1 | -7.807 | -7.902 | -7.931 |
| 3 | C8 | -7.895 | -7.910 | -7.910 |
| 4 | Ti4C4 | -7.919 | -7.869 | -7.869 |
| 5 | Zr4C2N1 | -7.782 | -7.770 | -7.776 |
| 6 | Zr4B1C5 | -7.761 | -7.728 | -7.748 |
| 7 | Zr2N2 | -7.431 | -7.591 | -7.592 |
| 8 | Zr3N1O4 | -7.567 | -7.501 | -7.565 |
| 9 | Zr4Sc1C5 | -7.572 | -7.481 | -7.510 |
| 10 | Zr4Be1C4 | -7.502 | -7.474 | -7.483 |
| 11 | Zr4C5F1 | -7.480 | -7.452 | -7.469 |
| 12 | Zr3O5F1 | -7.466 | -7.404 | -7.460 |
| 13 | Ti10C10 | -7.494 | -7.321 | -7.451 |
| 14 | Zr4Ti4C4 | -7.369 | -7.416 | -7.430 |
| 15 | Zr4B1O5 | -7.243 | -7.142 | -7.422 |
| 16 | Zr4Cu1C5 | -7.369 | -7.355 | -7.406 |
| 17 | Zr2B4 | -7.385 | -7.399 | -7.402 |
| 18 | Zr4Si1C5 | -7.510 | -7.341 | -7.397 |
| 19 | Ti5N4 | -7.454 | -7.383 | -7.389 |
| 20 | B2C10 | -7.581 | -7.359 | -7.364 |
| 21 | Y4O6 | -7.294 | -7.314 | -7.317 |
| 22 | Ti4C1O5 | -7.188 | -7.244 | -7.313 |
| 23 | Y1Zr4C5 | -7.335 | -7.236 | -7.274 |
| 24 | Zr4Al1C5 | -7.286 | -7.153 | -7.254 |
| 25 | Zr4H1C5 | -7.277 | -7.216 | -7.245 |
| 26 | Zr4B1N2 | -7.298 | -7.226 | -7.241 |
| 27 | Li1Zr4C5 | -7.224 | -7.217 | -7.231 |
| 28 | B2C5N1 | -7.257 | -7.177 | -7.228 |
| 29 | Zr4Sb1C5 | -7.206 | -7.063 | -7.200 |
| 30 | Ti6O10 | -7.277 | -7.152 | -7.198 |
| 31 | Zr4C5S1 | -7.413 | -7.174 | -7.192 |
| 32 | Zr4S1O5 | -7.114 | -6.990 | -7.156 |
| 33 | Ti4B1C5 | -7.085 | -6.959 | -7.153 |
| 34 | Zr4As1C2 | -7.274 | -7.115 | -7.146 |
| 35 | Zr4Ge1C5 | -7.222 | -6.916 | -7.144 |
| 36 | Y4C3O4 | -7.134 | -7.033 | -7.139 |
| 37 | Zr3Cl1O5 | -7.197 | -7.098 | -7.133 |
| 38 | Na1Zr4C5 | -7.114 | -7.095 | -7.108 |
| 39 | Zr4C5Br1 | -7.153 | -7.066 | -7.104 |
| 40 | Y1Zr4N5 | -7.083 | -7.032 | -7.058 |
| 41 | Zr4Te1C5 | -7.194 | -7.053 | -7.058 |
| 42 | Sc5O7 | -7.183 | -7.020 | -7.043 |
| 43 | Y4O5F1 | -7.080 | -7.028 | -7.043 |
| 44 | C9N1 | -7.283 | -6.947 | -7.001 |
| 45 | Zr4Ga1C5 | -7.092 | -6.794 | -6.927 |
| 46 | Mg1Zr4C5 | -7.078 | -6.901 | -6.916 |
| 47 | Ti3B4 | -7.129 | -6.853 | -6.866 |
| 48 | Zr4Se1C5 | -7.135 | -6.748 | -6.835 |
| 49 | Zr4P1C5 | -7.137 | -6.743 | -6.830 |
| 50 | Y1B1C5 | -7.140 | -6.699 | -6.819 |

Table S2. The formula between the number of local environments and the number of elements in a tetragonal structure.

| $N_{elem}$ | $N_{descriptor}$ |
|---|---|
| 1 | 1 |
| 2 | $C_2^1 * (C_2^1 + C_2^2) = 6$ |
| 3 | $C_3^1 * (C_3^1 + C_3^2 + C_3^3) = 21$ |
| 4 | $C_4^1 * (C_4^1 + C_4^2 + C_4^3 + C_4^4) = 60$ |
| ... | ... |
| $n$ | $C_n^1 * (C_n^1 + C_n^2 + C_n^3 + C_n^4) = \dfrac{n^2(n+1)(n^2 - 3n + 14)}{24}$ |

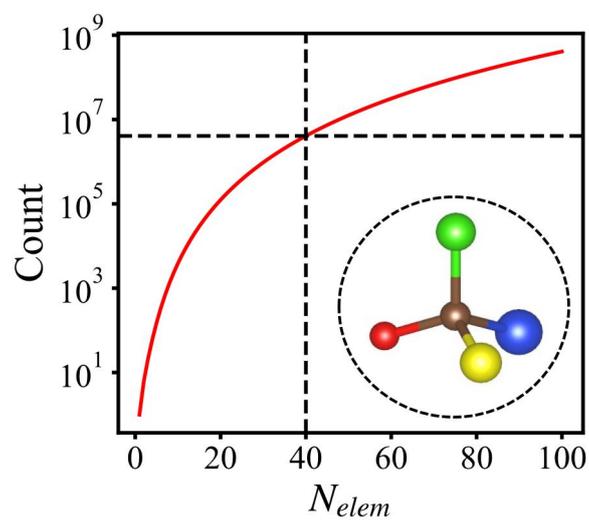

Fig. S2. The relationship between the number of local environments and the number of elements in a tetragonal structure.

Fig. S3. The 42 elements that we have selected are marked in green.

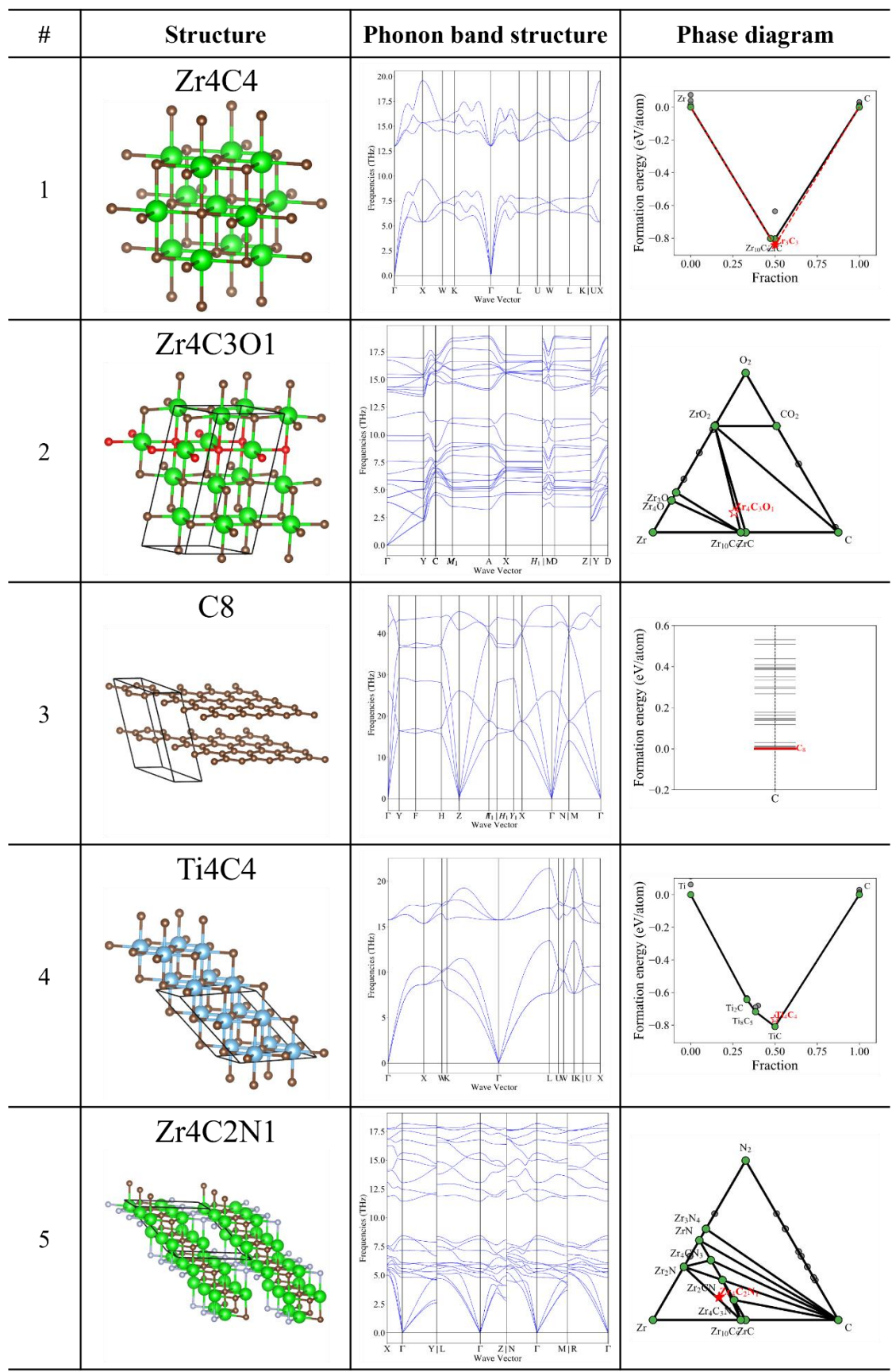

Fig. S4. The structures, phonon band structures and phase diagrams of $Zr_4C_4$, $Zr_4C_3O_1$, $C_8$, $Ti_4C_4$ and $Zr_4C_2N_1$, respectively.

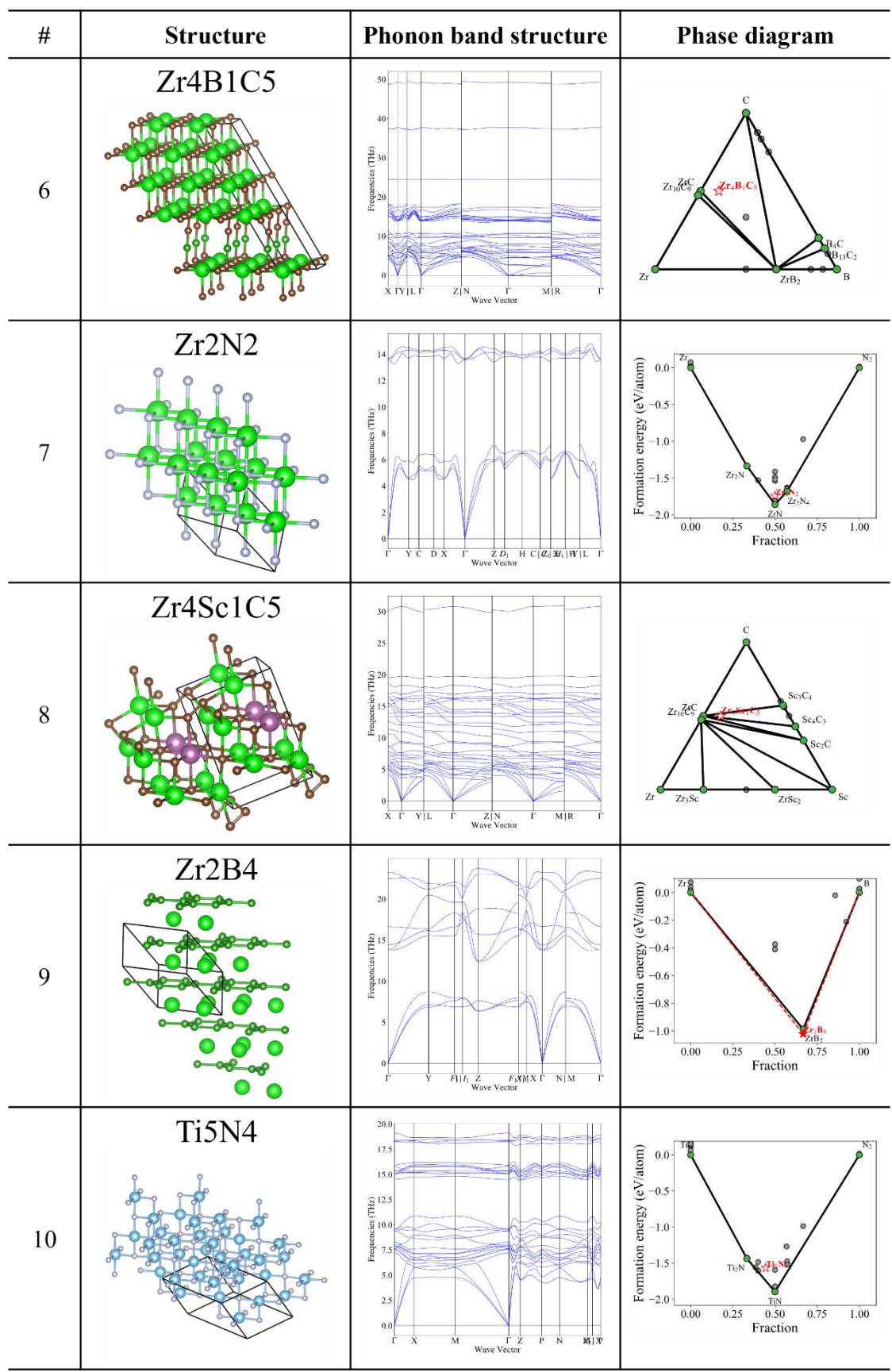

Fig. S5. The structures, phonon band structures and phase diagrams of $Zr_4B_1C_5$, $Zr_2N_2$, $Zr_4Sc_1C_5$, $Zr_2B_4$ and $Ti_5N_4$, respectively.